\documentclass[letterpaper, 10 pt, conference]{ieeeconf}  

\usepackage{marvosym}  
\usepackage{url}
\usepackage{hyperref}

\IEEEoverridecommandlockouts                              

\title{\LARGE \bf
EGSTalker: Real-Time Audio-Driven Talking Head Generation with Efficient Gaussian Deformation
}

\author{
Tianheng Zhu$^1$, Yinfeng Yu$^{1}$$^{,\mbox{\Letter}}$, Liejun Wang$^1$, Fuchun Sun$^2$, and Wendong Zheng$^3$%
\thanks{$^{\mbox{\Letter}}$\small Yinfeng Yu is the corresponding author (e-mail: yuyinfeng@xju.edu.cn).}%
\thanks{\textsuperscript{*}Project page and code: \url{https://github.com/ZhuTianheng/EGSTalker.}}%
\\
$^1$Xinjiang Multimodal Intelligent Processing and Information Security Engineering Technology Research Center,\\
School of Computer Science and Technology, Xinjiang University, Urumqi 830017, China
\\
$^2$Department of Computer Science and Technology, Tsinghua University, Beijing 100091, China
\\
$^3$School of Electrical Engineering and Automation, Tianjin University of Technology, Tianjin 300382, China%
}

\usepackage{multirow}
\usepackage{graphicx}
\usepackage{amsmath}
\usepackage{amssymb}  
\usepackage{mathptmx}  
\usepackage{booktabs}
\usepackage{bm} 
\usepackage{cite}

\begin{document}

\maketitle

\thispagestyle{empty}
\pagestyle{empty}


\begin{abstract}

This paper presents EGSTalker, a real-time audio-driven talking head generation framework based on 3D Gaussian Splatting (3DGS). Designed to enhance both speed and visual fidelity, EGSTalker requires only 3–5 minutes of training video to synthesize high-quality facial animations. The framework comprises two key stages: static Gaussian initialization and audio-driven deformation. In the first stage, a multi-resolution hash triplane and a Kolmogorov-Arnold Network (KAN) are used to extract spatial features and construct a compact 3D Gaussian representation. In the second stage, we propose an Efficient Spatial-Audio Attention (ESAA) module to fuse audio and spatial cues, while KAN predicts the corresponding Gaussian deformations. Extensive experiments demonstrate that EGSTalker achieves rendering quality and lip-sync accuracy comparable to state-of-the-art methods, while significantly outperforming them in inference speed. These results highlight EGSTalker’s potential for real-time multimedia applications.

\end{abstract}

\section{INTRODUCTION}

Audio-driven talking head video generation has emerged as a significant research topic in computer vision and graphics, with diverse applications in digital humans, virtual reality, and video conferencing \cite{wav2lip,synobama,iplap,SAAVN,YinfengIJCAI2023MACMA,fsaavn,ttt}. Despite considerable progress, existing approaches still face challenges regarding synthesis quality and computational efficiency.

Recent methods \cite{guo2021ad,tang2022real,ernerf} primarily employ Neural Radiance Fields (NeRF) \cite{nerf}, leveraging implicit representations via deep neural networks to achieve high-quality dynamic talking head synthesis. Although these methods deliver photorealistic results, they often encounter issues like poor lip-audio synchronization, high computational overhead, extensive training durations, and slow rendering speeds. 

\begin{figure}[thpb]
	\centering
	\includegraphics[scale=0.24]{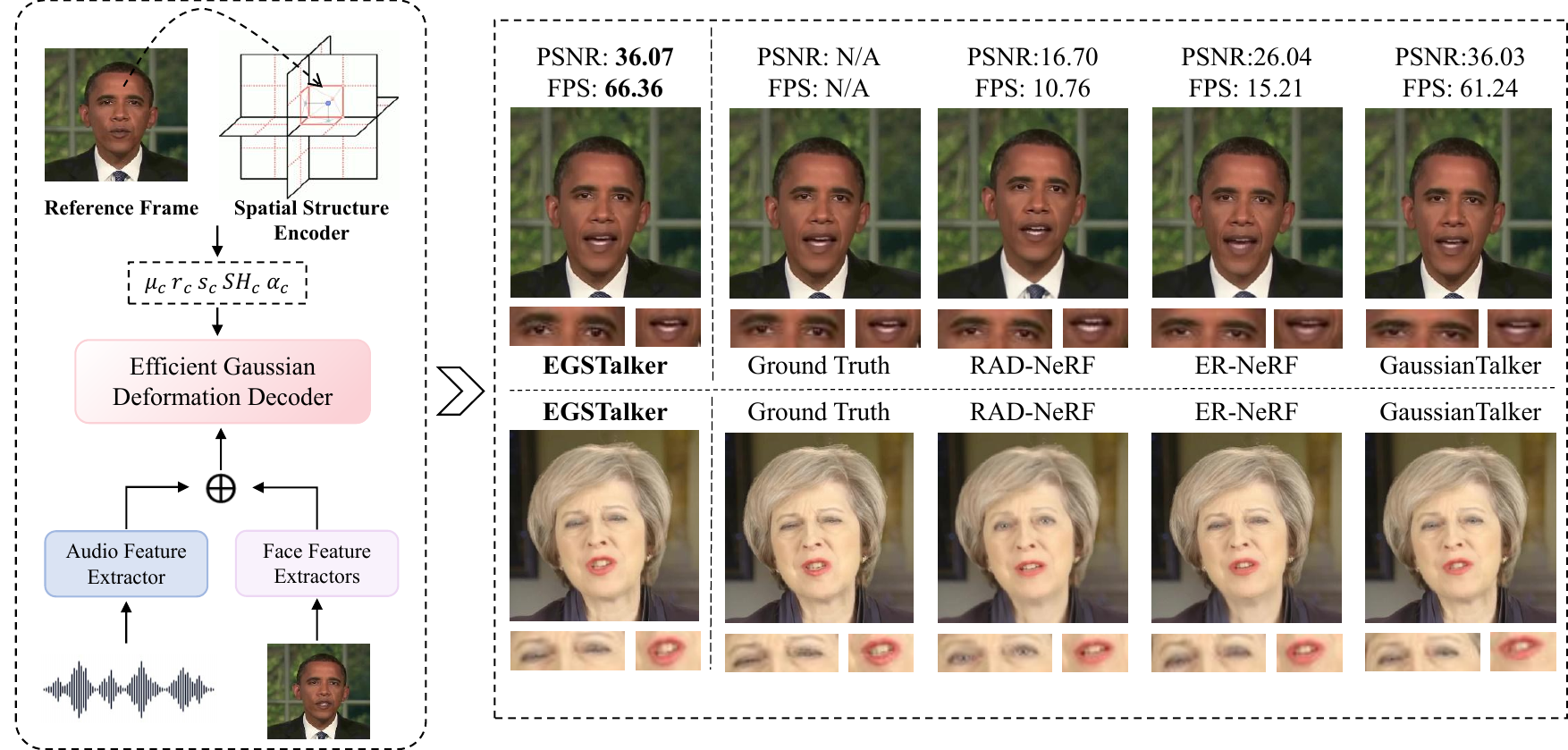}  
	\caption{EGSTalker accelerates talking head synthesis by combining a structured head representation with an optimized spatial-audio attention mechanism, enabling efficient 3D Gaussian deformation with improved clarity and motion fidelity.}
	\label{headfig}
\end{figure}

Alternatively, 3D Gaussian Splatting (3DGS) \cite{3dgs} has become a promising approach due to its rapid rendering capabilities without compromising visual quality. However, existing 3DGS-based methods integrating audio \cite{li2025talkinggaussian,cho2024gaussiantalker} typically rely on lightweight multilayer perceptrons (MLPs) or conventional attention mechanisms, limiting both fusion quality and computational efficiency.

To address these limitations, we propose EGSTalker, an innovative framework for efficient and high-fidelity audio-driven talking head video generation. EGSTalker leverages 3DGS to model head structures and facial expressions, introducing a spatial structure encoder and an efficient Gaussian deformation decoder. The encoder combines a multi-resolution hash triplane \cite{hexplane} with a Kolmogorov-Arnold Network (KAN) \cite{kan} to hierarchically encode the head region and robustly map spatial features to Gaussian parameters. The decoder integrates a spatial-audio attention (ESAA) module and periodic positional encoding (PPE) \cite{fan2022faceformer}, using an agent-based mechanism \cite{han2025agent} to effectively fuse audio and spatial features while maintaining low computational complexity and ensuring synchronized facial dynamics.

The main contributions of this paper are as follows:
\begin{itemize}
	\item We propose a spatial feature encoder that encodes head spatial regions into Gaussian parameters, enabling more precise representation of spatial point relationships.
	\item We design an efficient Gaussian deformation decoder that rapidly fuses audio and spatial features while predicting Gaussian deformations to generate audio-synchronized frames. 
	\item Experimental results demonstrate that EGSTalker achieves significant advantages in facial fidelity, audio-video synchronization quality, and rendering efficiency, exhibiting strong practical value.
\end{itemize}

\section{METHODS}

\begin{figure*}[t]
	\centering
	\includegraphics[width=0.8\textwidth]{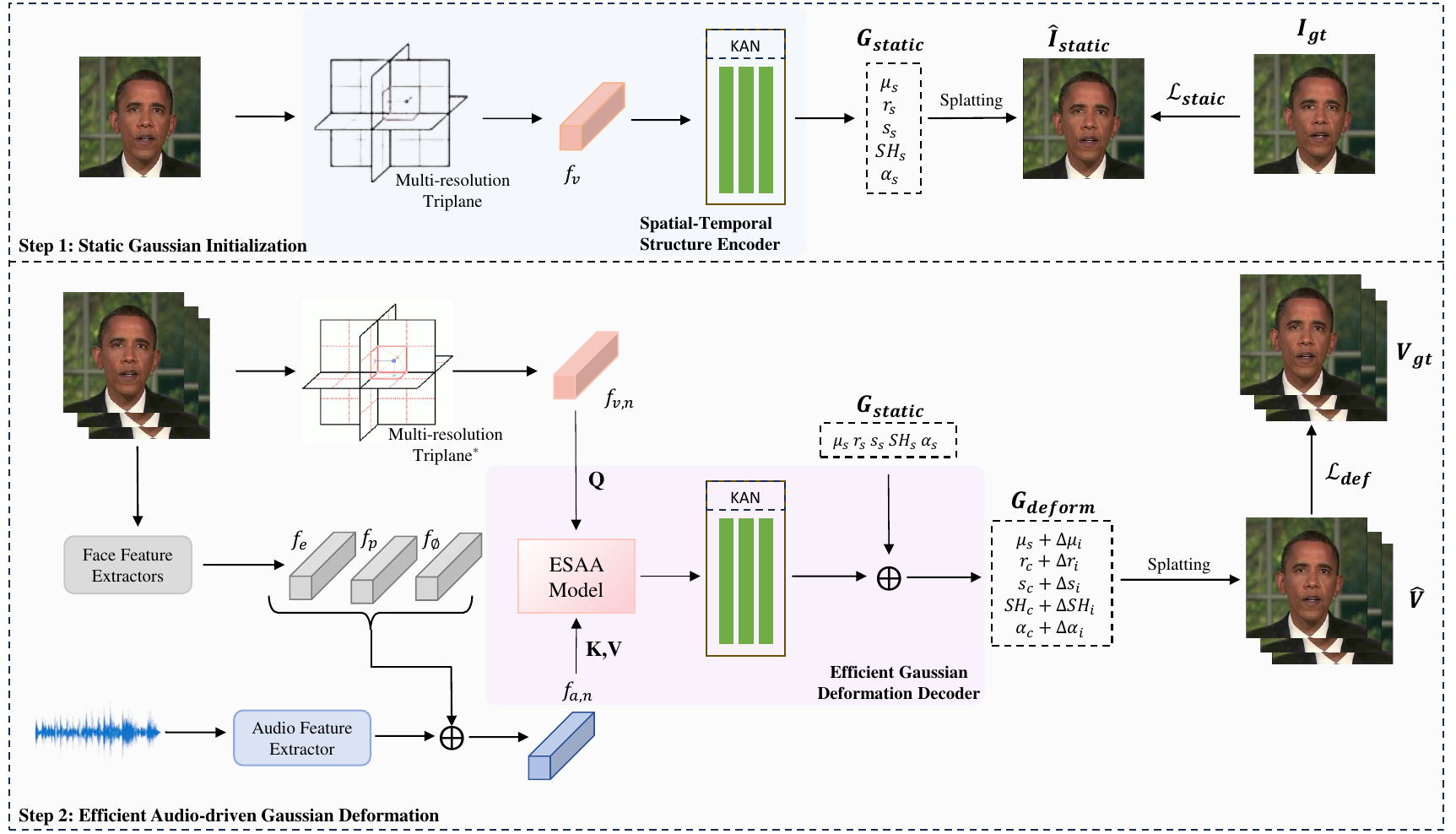}
	\caption{Overview of the EGSTalker framework. The model employs a two-stage training strategy: the first stage constructs static 3D Gaussian representations, while the second stage uses an audio-guided deformation decoder to predict dynamic facial motions.}
	\label{fig:framework}
\end{figure*}

\subsection{Overview}
\label{4.1}
The EGSTalker model is designed for efficient and high-fidelity talking head video generation using limited training data. As illustrated in Fig.~\ref{fig:framework}, the framework follows a two-stage training strategy.

In the first stage, a spatial structure encoder extracts spatial features from video frames to construct a static 3D Gaussian representation (3DGS) of the subject's head (Section~\ref{4.2}). In the second stage, audio features are extracted alongside facial attributes (e.g., eye blinking) and fused with the spatial features. A Gaussian deformation decoder then predicts dynamic Gaussian deformations, generating audio-driven facial motion (Section~\ref{4.3}). The loss functions utilized in both stages are discussed in Section~\ref{4.4}.

\subsection{Static Gaussian Initialization}
\label{4.2}

This section details the construction of a static 3D Gaussian representation for the subject's head in the training video. Facial motions exhibit regional characteristics, where spatially adjacent points follow similar motion trajectories, while different regions display distinct movement patterns. Although 3DGS can effectively model a static head, it struggles to capture both intra- and inter-region spatial relationships. To address this, we introduce a spatial structure encoder that extracts spatial features and predicts 3D Gaussian parameters.

As shown in Fig.~\ref{fig:framework}, the spatial structure encoder consists of two key components: (1) a multi-resolution hash triplane encoder for structured spatial feature extraction and (2) a Kolmogorov-Arnold Network (KAN) for efficient mapping to the 3D Gaussian space. Hash triplane encoder hierarchically encodes the head region using three orthogonal 2D grids, reducing computational redundancy in sparse regions while preserving fine details in dense areas. Each spatial point \( \mathbf{x} = (x, y, z) \) is projected onto these grids, interpolated across resolutions, and fused via the Hadamard product to generate a compact feature vector \( f_{v}(\mathbf{x}) \).

Traditional MLP-based mappings suffer from high-dimensional nonlinearity, leading to information loss and computational inefficiency. Instead, we employ KAN, which leverages polynomial approximation to accurately capture complex relationships between facial features and 3D Gaussian parameters.S The final mapping is formulated as:
\begin{equation}
	\mathcal{G}_{static} \{ \mu_{s}, s_{s}, r_{s}, SH_{s}, \alpha_{s} \} =KAN(f_{v}(\mathbf{x})).
	\label{kan1}
\end{equation}
This structured representation provides a robust foundation for subsequent audio-driven deformations.

\subsection{Efficient Audio-driven Gaussian Deformation}
\label{4.3}
To achieve audio-synchronized talking head generation, static Gaussians must be dynamically deformed based on audio input. We design an efficient Gaussian deformation decoder to capture the mapping from audio signals to geometric transformations, as illustrated in Fig.~\ref{fig:esaa}.

Prior works \cite{ernerf} demonstrate that explicit cross-modal attention surpasses MLP-based methods in modeling facial dynamics. Unlike approaches that directly fuse spatial and audio features via MLPs \cite{guo2021ad}, we introduce the Efficient Spatial-Audio Attention (ESAA) module to enhance audio-spatial interactions.

Existing methods commonly adopt element-wise modulation to integrate audio into spatial representations \cite{ernerf, li2025talkinggaussian}, under the assumption that static 3D coordinates correspond to consistent facial regions. This assumption is often invalid due to dynamic facial variations. Although \cite{cho2024gaussiantalker} address this via a spatial-audio attention module with cross-attention, the high computational overhead hinders real-time performance.

To address this, our ESAA module employs a lightweight yet expressive attention mechanism, capturing global audio-spatial dependencies with reduced complexity and enabling real-time inference with improved audio-motion alignment.

\subsubsection{Efficient Spatial-Audio Attention}

\begin{figure}[h]
	\centering
	\includegraphics[width=\linewidth]{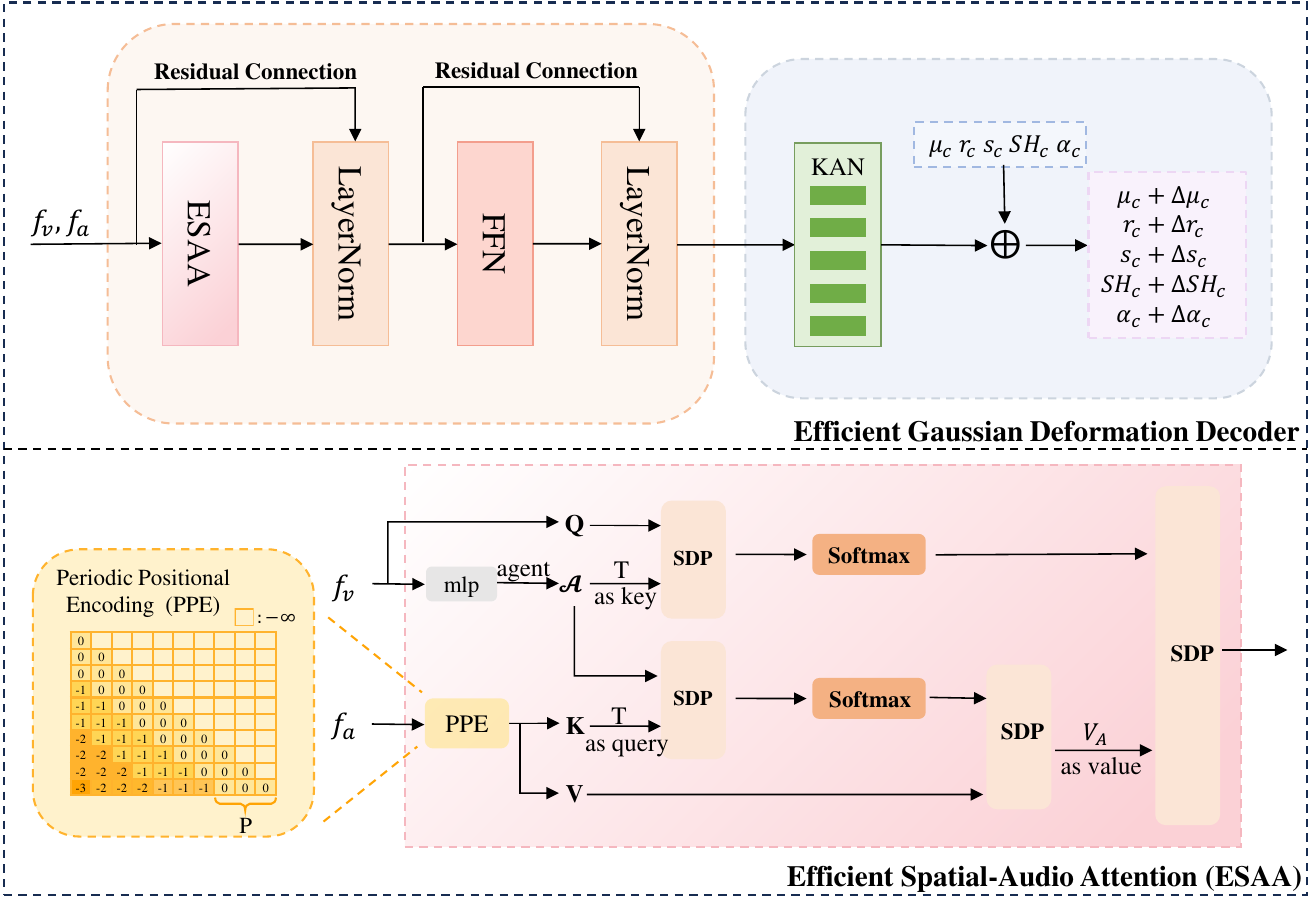}
	\caption{Overview of the Efficient Gaussian Deformation Decoder (a) and the ESAA module (b). The decoder predicts Gaussian attribute offsets for audio-driven deformation, with ESAA enabling spatial-audio interaction and PPE encoding temporal information.}
	\label{fig:esaa}
\end{figure}

Inspired by \cite{han2025agent}, we design the ESAA module to efficiently fuse spatial and audio features using agent tokens \( \mathcal{A} \), as shown in Fig.~\ref{fig:esaa}. A lightweight MLP projects spatial features to agent tokens: \( \mathcal{A} = \mathbf{MLP}(f_v) \).

These tokens act as Query in an initial attention step to aggregate proxy audio features \( V_{\mathcal{A}} \). Then, \( \mathcal{A} \) serves as Key, \( V_{\mathcal{A}} \) as Value, and the original spatial features \( f_v \) as Query, enabling global broadcast across the spatial domain:
\begin{equation}
	ACA\left(f_v, f_a\right) = \underbrace{\mathrm{SDP}\left(f_v, \mathcal{A}, \underbrace{\mathrm{SDP}(\mathcal{A}, f_a, f_a)}_{\text{Agent Aggregation}}\right)}_{\text{Agent Broadcast}},
\end{equation}
where the scaled dot-product attention mechanism is defined as: $\mathrm{SDP}(Q, K, V) = \mathrm{Softmax}\left(\frac{QK^T}{\sqrt{d_k}}\right)V.$

To reduce cost, the agent token count is restricted to 1\% of the spatial sequence length, reducing complexity from \( O(N^2 \cdot d) \) to \( O(N \cdot n \cdot d) \), where \( N \), \( n \), and \( d \) denote spatial length, agent token count, and feature dimension.

To encode the periodicity of audio-driven motion, we adopt Periodic Positional Encoding (PPE)~\cite{fan2022faceformer}:
$\mathrm{PPE}(t, 2i) = \sin\left(\frac{t \bmod p}{10000^{2i/d}}\right)$, 
$\mathrm{PPE}(t, 2i+1) = \cos\left(\frac{t \bmod p}{10000^{2i/d}}\right)$, 
where $t$ is the timestep, $d$ the feature dimension, and $p$ the period.

Beyond audio, non-audio factors such as blinking and camera pose affect facial motion. To address this, we incorporate eye features \( f_e \), pose embeddings \( f_p \), and a global zero-vector \( f_{\emptyset} \), all projected into a shared latent space and concatenated with \( f_a \). The fused representation is then integrated with spatial features by ESAA to synthesize temporally coherent and expressive facial motion.

\subsubsection{Gaussian Deformation Prediction}

Following \cite{4dgs}, we model temporal dynamics by predicting frame-wise offsets to Gaussian attributes. Instead of using MLPs as in \cite{4dgs}, we employ KAN for its superior capacity to capture complex nonlinear mappings. Given the audio-aware spatial feature \( f_d \), KAN predicts the deformation offsets:
\begin{equation}
	\mathcal{G}_{\text{deform}}\{\Delta \mu_n, \Delta s_n, \Delta r_n, \Delta SH_n, \Delta \alpha_n\} = KAN(f_d(\mathbf{x}_n)).
	\label{kan2}
\end{equation}
Here, \( \mathbf{x}_n \) denotes a spatial location in frame \( n \), and the offsets correspond to adjustments in position, scale, rotation, spherical harmonics, and opacity.

The final video sequence \( \hat{V} \) is obtained by splatting the combined static and dynamic Gaussian parameters for each frame:
\begin{equation}
	\hat{V} = \{ I_i \} = \text{Splatting}(\mathcal{G}_{\text{static}} + \mathcal{G}_{\text{deform}}),
	\label{video1}
\end{equation}
where \( I_i \) denotes the \( i \)-th rendered frame.

\subsection{Training Objectives}
\label{4.4}

We adopt a two-stage training strategy aligned with our Gaussian-based pipeline. In the static initialization stage, we follow the 3DGS framework \cite{3dgs} and optimize a composite loss comprising $L1$ color loss $L_{\mathrm{L1}}$, D-SSIM loss $L_{\mathrm{D-SSIM}}$, and LPIPS loss \cite{zhang2021flow} to enhance perceptual fidelity:
\begin{equation}
	L_{\mathrm{static}} = L_{\mathrm{L1}} + \lambda_{\mathrm{D-SSIM}} L_{\mathrm{D-SSIM}} + \lambda_{\mathrm{lpips}} L_{\mathrm{lpips}},
\end{equation}
where $\lambda_{\mathrm{D-SSIM}}$ and $\lambda_{\mathrm{lpips}}$ balance structural similarity and perceptual quality.

In the Gaussian deformation stage, we build upon the static loss by introducing a lip reconstruction loss $L_{\mathrm{lip}}$ \cite{cho2024gaussiantalker}, computed from cropped lip regions using facial landmarks \cite{bulat2017far}, to improve synchronization:
\begin{equation}
	L_{\mathrm{def}} = L_{\mathrm{L1}} + \lambda_{\mathrm{D-SSIM}} L_{\mathrm{D-SSIM}} + \lambda_{\mathrm{lpips}} L_{\mathrm{lpips}} + \lambda_{\mathrm{lip}} L_{\mathrm{lip}},
\end{equation}
where all $\lambda$ terms denote corresponding loss weights.

\section{Experiment}

\begin{figure*}[t]
	\centering
	\includegraphics[width=0.75\textwidth]{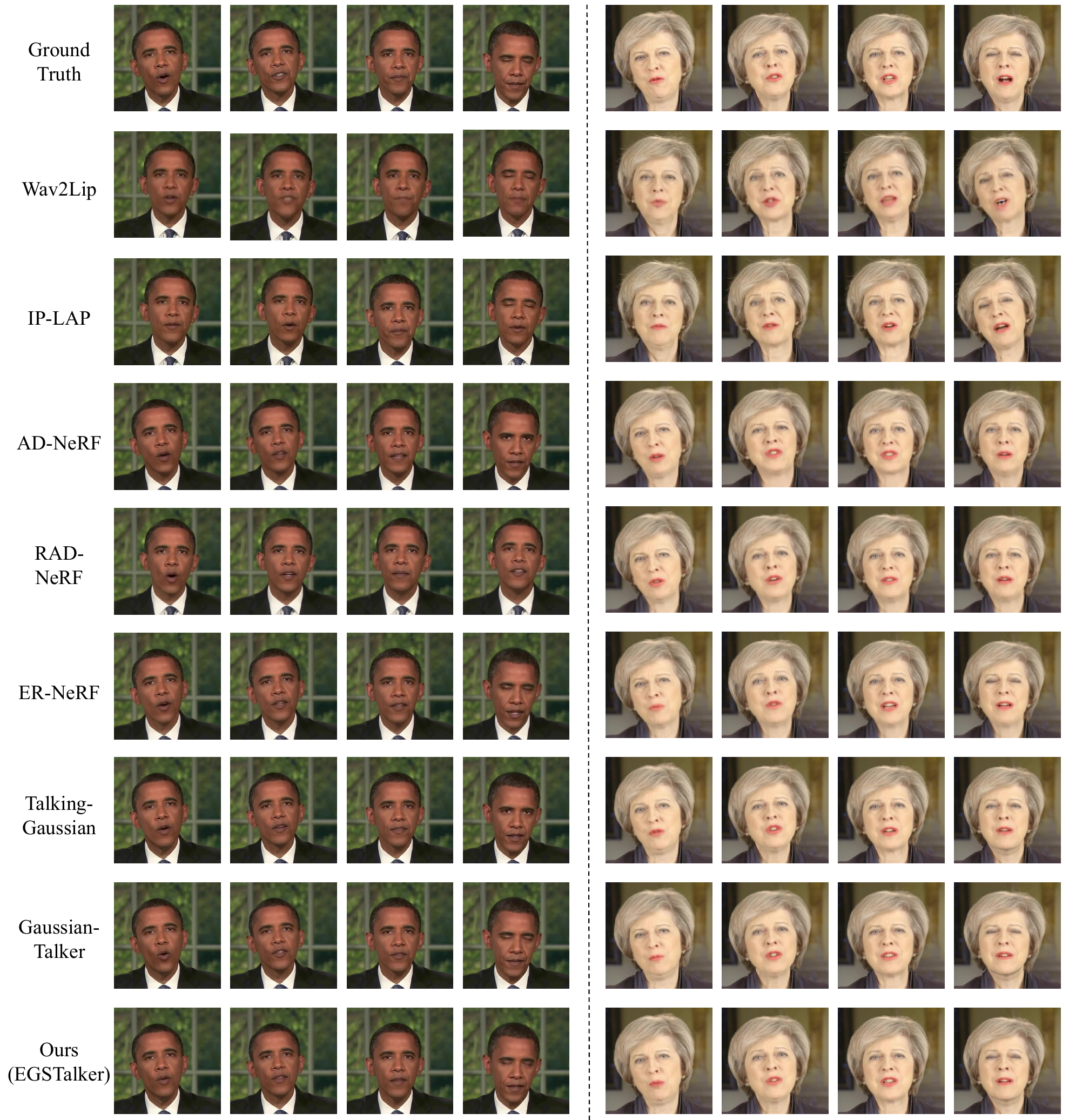}
	\caption{Qualitative results of the self-driven setting on the Obama and May datasets.Our method achieves competitive results in reconstruction quality and lip synchronization compared to the state-of-the-art 3DGS-based method, GaussianTalker, excelling in head pose and facial expression control.}
	\label{fig:result-fig}
\end{figure*}

\subsection{Experimental Settings}

\begin{figure}[h]
	\centering
	\includegraphics[width=0.85\linewidth]{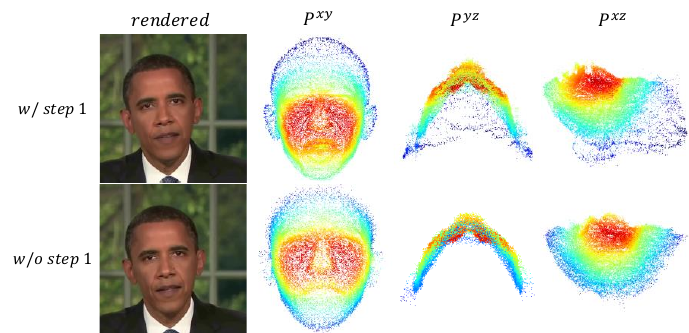}
	\caption{Effect of Static Gaussian Initialization on 3D point distribution. Without initialization, key facial regions (e.g., lips, eyes) exhibit sparse point density, degrading expression modeling. Initialization yields denser, more structured distributions, enhancing reconstruction and dynamic fidelity.}
	\label{fig:mesh}
\end{figure}
\begin{figure}[h]
	\centering
	\includegraphics[width=0.9\linewidth]{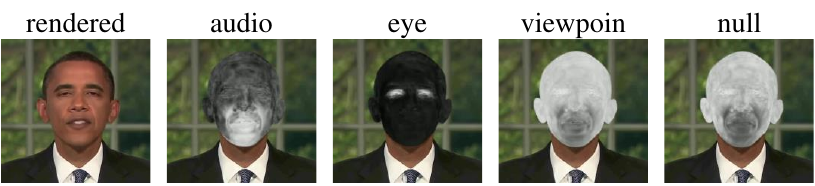}
	\caption{Visualization of attention maps from the ESAA module. From left to right: rendered image, attention to lip synchronization, eye blinks, head orientation, and temporal consistency.}
	\label{fig:att}
\end{figure}

\subsubsection{Dataset and Pre-processing}

We construct our dataset using four high-resolution video clips sourced from prior works \cite{guo2021ad,ernerf,tang2022real}, featuring three male and one female subject. Each clip contains approximately 6500 frames at 25 FPS. The portrait frames are center-cropped and resized to $512 \times 512$, except for the ``Obama'' clip, which is resized to $450 \times 450$. We split the dataset into training and testing subsets. All experiments are conducted on a single NVIDIA Tesla T4 GPU (16GB VRAM).

\subsubsection{Comparison Baselines}

We compare our EGSTalker framework against several representative baselines. These include NeRF-based models such as AD-NeRF \cite{guo2021ad}, RADNeRF \cite{tang2022real}, ER-NeRF \cite{ernerf}, and the 3DGS-based GaussianTalker \cite{cho2024gaussiantalker}, which leverage identity-specific radiance fields and 3D point clouds conditioned on speech-video pairs. Additionally, we include comparisons with state-of-the-art 2D methods like Wav2Lip \cite{wav2lip} and IP-LAP \cite{iplap} for a broader evaluation.
\begin{table*}
	\caption{The quantitative results of the \textbf{self-driven} setting. The best and second-best methods are in \textbf{bold} and \underline{underline}, respectively. Our EGSTalker achieves competitive results while significantly improving inference efficiency.
	}
	\label{tab:db}
	\centering
	\begin{tabular}{ccccccccccc}
		\toprule
		\multirow{2}{*}{\textbf{Methods}} & \multicolumn{4}{c}{\textbf{Rendering Quality}} & \multicolumn{3}{c}{\textbf{Motion Quality}} & \multicolumn{2}{c}{\textbf{Efficiency}}          \\  
		& PSNR$\uparrow$     & SSIM$\uparrow$     & LPIPS$\downarrow$   & FID$\downarrow$       & LMD$\downarrow$     & LSE-D$\downarrow$   & LSE-C$\uparrow$             & Time$\downarrow$       & \multicolumn{1}{c}{FPS$\uparrow$}          \\ \midrule
		GT & N/A &1   &0   &0   &0   & 6.745   & 8.491      & N/A & \multicolumn{1}{c}{N/A} \\
		Wav2Lip~\cite{wav2lip}    & 16.507   & 0.847    & 0.2994  & 27.942         & 4.978   & \textbf{7.019}   & \textbf{7.972}      & -            & \multicolumn{1}{c}{8.759}        \\
		Ip-LAP~\cite{iplap}      & 16.418   & 0.841    & 0.2992  & 29.853         & 4.665   & 8.99    & 5.256      & -            & \multicolumn{1}{c}{1.314}        \\
		AD-NeRF~\cite{guo2021ad}     & 25.794   & 0.9643   & 0.0842  & 18.289         & 2.932   & 9.839   & 5.105      & 167.6h      & \multicolumn{1}{c}{0.04}         \\
		RAD-NeRF~\cite{tang2022real}    & 16.706   & 0.855    & 0.281   & 22.209         & 3.291   & 8.669   & 6.288      & 13.8h       & \multicolumn{1}{c}{10.76}        \\
		ER-NeRF~\cite{ernerf}      & 26.047   & 0.961    & 0.0635  & 7.637          & 2.547   & \underline{7.913}   & \underline{7.054}      & 8.9h        & \multicolumn{1}{c}{15.21}        \\
        TalkingGaussian~\cite{li2025talkinggaussian}   & 35.21   & 0.990    & 0.0189  & 3.398          & \underline{2.538}   & 8.051   & 6.963      & \textbf{1.5h}        & \multicolumn{1}{c}{\textbf{70.42}}        \\
		GaussianTalker~\cite{cho2024gaussiantalker}    & \underline{36.034}   & \underline{0.992}    & \underline{0.0224}  & \underline{2.431}          & 2.614   & 8.274   & 6.964     & 4.5h        & \multicolumn{1}{c}{59.24}        \\
		Ours(EGStalker)    & \textbf{36.070}  & \textbf{0.992} & \textbf{0.0223} & \textbf{2.424}    & \textbf{2.536}   & 8.237   & 6.966      & \underline{3.7h}        & \multicolumn{1}{c}{\underline{68.51}} \\ \bottomrule
	\end{tabular}
\end{table*}

\subsection{Quantitative Evaluation}

\subsubsection{Comparison Settings and Metrics}

Following previous works \cite{guo2021ad,ernerf,cho2024gaussiantalker}, our comparison consists of two different settings: self-driven and cross-driven. In the self-driven setting, we evaluate the accuracy of head reconstruction for specific identities using the test subset. Several reconstruction metrics are employed to assess image quality, including Peak Signal-to-Noise Ratio (\textbf{PSNR}), Structural Similarity Index (\textbf{SSIM}) \cite{SSIM}, and Learned Perceptual Image Patch Similarity (\textbf{LPIPS}) \cite{LPIPS}. It is noteworthy that these metrics are measured on the facial region. Additionally, we use Fréchet Inception Distance (\textbf{FID}) \cite{fid} to evaluate the realism of the reconstructed faces. To assess the synchronization of audio-lip movements, we use Landmark Distance (\textbf{LMD}), Lip Sync Error - Distance (\textbf{LSE-D}), and Lip Sync Error - Confidence (\textbf{LSE-C}) \cite{wav2lip}. Finally, we compare the training time (\textbf{Time}) and frames per second (\textbf{FPS}) as metrics to evaluate the efficiency and generation speed of each method.

For the cross-driven setting, all methods are driven by completely unrelated audio tracks to evaluate lip synchronization. The audio clips used in this setting are extracted from the SynObama demo\cite{synobama}. Due to the absence of ground-truth images, we only use \textbf{LMD}, \textbf{LSE-C} and \textbf{LSE-D} to assess the synchronization accuracy of audio-lip movements.

\subsubsection{Self-driven Evaluation}

As shown in Table~\ref{tab:db}, Wav2Lip and IP-LAP achieve strong lip synchronization but suffer from limited visual fidelity. NeRF-based methods \cite{guo2021ad,tang2022real,ernerf} offer higher image quality but are computationally expensive. GaussianTalker strikes a balance between quality and speed but is not yet real-time. EGSTalker surpasses all baselines in both rendering quality and synchronization accuracy while achieving the highest FPS, making it ideal for real-time, high-fidelity talking head generation.

\subsubsection{Cross-driven Evaluation}
Table~\ref{tab:cross-driven} highlights our method's robustness under cross-modal scenarios, where the input audio is unrelated to the target identity. Compared to NeRF-based methods, EGSTalker achieves competitive or superior synchronization performance, matching or outperforming state-of-the-art models such as GaussianTalker, demonstrating strong generalization and natural facial dynamics under unseen conditions.

\begin{table}
	\centering 
	\caption{The quantitative results of the \textbf{cross-driven} setting. We extract two audio clips from SynObama demo \cite{synobama} to drive each method and compare lip synchronization}
	\label{tab:cross-driven}
	\begin{tabular}{c@{\hskip 4pt}c@{\hskip 4pt}c@{\hskip 4pt}c@{\hskip 4pt} c@{\hskip 4pt}c@{\hskip 4pt}c}  
		\toprule
		\multirow{2}{*}{\textbf{Methods}} & \multicolumn{3}{c}{\textbf{Testset A}}           & \multicolumn{3}{c}{\textbf{Testset B}}           \\  
		& LMD$\downarrow$ & LSE-C$\uparrow$ & LSE-D$\downarrow$  & LMD$\downarrow$ & LSE-C$\uparrow$    & LSE-D$\downarrow$            \\ 
		\midrule
		AD-NeRF~\cite{guo2021ad} & 7.716 & 4.932 & 10.547 & 8.379 & 4.443 & 10.707          \\
		RAD-NeRF~\cite{tang2022real} & 7.575 & 6.697 & 8.665 & 8.562 & \underline{6.669} & \underline{8.620} \\
		ER-NeRF~\cite{ernerf} & 7.458 & \underline{6.865} & \textbf{8.361} & 8.362 & \textbf{7.061} & \textbf{8.269} \\
        TalkingGaussian~\cite{li2025talkinggaussian} & \textbf{7.450} & 6.136 & 9.265 & \underline{8.323} & 6.381 & 8.637 \\
		GaussianTalker~\cite{cho2024gaussiantalker} & 7.994 & 6.260 & 9.523 & 8.687 & 6.618 & 8.950 \\
		Ours(EGStalker) & \underline{7.459} &\textbf{6.945} & \underline{8.470} & \textbf{8.224} & 6.461 & 8.862 \\
		\bottomrule
	\end{tabular}
\end{table}

\subsection{Qualitative Evaluation}

Fig.~\ref{fig:result-fig} presents qualitative comparisons under the self-driven setting using samples from the Obama and May datasets. Four representative frames are selected to assess reconstruction fidelity and lip synchronization.

2D-based methods such as Wav2Lip and IP-LAP excel at lip synchronization but struggle with detailed facial representation and are sensitive to head pose variations. NeRF-based models like AD-NeRF and RAD-NeRF lack explicit control over eye dynamics, leading to less realistic expressions. In contrast, our method maintains consistent head pose, models fine-grained facial expressions (including eye blinks), and achieves high lip synchronization accuracy. Compared to the 3DGS-based GaussianTalker, our approach offers enhanced control over dynamic facial attributes while preserving high-fidelity reconstruction.

\subsection{Ablation Study}
In this section, we provide ablation studies to validate the efficacy of the design choices of our model.

\subsubsection{Impact of Key Components}

We conduct an ablation study to evaluate the contribution of core components, including KAN, ESAA, PPE, and Step 1 (Static Gaussian Initialization). As shown in Table~\ref{tab:xr-kan}, the complete model achieves the best performance.
Replacing the MLP with KAN significantly enhances reconstruction quality, confirming its effectiveness in modeling complex nonlinear mappings. ESAA and PPE jointly improve lip synchronization, with PPE notably strengthening temporal coherence. Removing Step 1 leads to a clear drop in reconstruction accuracy, underscoring its role in establishing a high-quality Gaussian distribution.Fig.~\ref{fig:mesh} further illustrates the impact of Step 1. Without static initialization, the point density around key facial regions (e.g., lips, eyes) is visibly reduced, compromising the model's ability to represent fine-scale dynamics. Incorporating Step 1 yields denser and more structured Gaussians in expression-critical areas, thereby improving both geometric precision and audiovisual alignment.
\begin{table}
	\caption{Ablation study of our contributions under the self-reconstruction setting.}
	\label{tab:xr-kan}
	\begin{tabular}{ccccc}
		\toprule
		Methods   & PSNR$\uparrow$  & LPIPS$\downarrow$   & LMD$\downarrow$   &LSE-C$\uparrow$    \\ \midrule
		w/o KAN, ESAA, ppe & 36.034          & 0.0224          & 2.614              & \underline{6.964}    \\
		w/o ESAA, PPE     & \textbf{36.415} & \textbf{0.0218}   & 2.638              & 6.43           \\
		w/o PPE           & \underline{36.104}    & \underline{ 0.0218}    & \underline{2.537}   &6.851          \\
		w/o Step 1         & 35.865          & 0.0248          & 2.651                   & 6.461          \\
		All(Ours)         & 36.07           & 0.0223          & \textbf{2.536}  & \textbf{6.966} \\ \bottomrule
	\end{tabular}
\end{table}

\begin{table}
	\caption{Ablation study on the impact of the number of agent tokens in the ESSA module.}
	\label{tab:xr-agent}
	\begin{tabular}{ccccccc}
		\toprule
		Methods        & Num    & LMD$\downarrow$      & LSE-D$\downarrow$    & LSE-C$\uparrow$   & AUE$\downarrow$     & FPS$\uparrow$ \\ \midrule
		w/o ESAA       & -      & 2.559    & 8.571   & 6.754    & 0.761   & 57.4       \\ \midrule
		\multirow{4}{*}{w/ ESAA} 
		& 0.16\%           & 2.61         & 8.335          & 6.883          & 0.661        & \textbf{74.1}       \\
		& 0.25\%          & 2.55         & 8.479          & 6.84           & \underline{0.611}        & \underline{69.1}       \\
		& 0.5\%          & \underline{2.536}        & \textbf{8.237}          & \textbf{6.966}          & 0.793        & 68.5        \\
		& 1.0\%          & \textbf{2.508}        & \underline{8.319}          & \underline{6.877}           & \textbf{0.606}        & 65.6       \\ \bottomrule
	\end{tabular}
\end{table}

\subsubsection{Effect of Agent Tokens in the ESAA Module}
We investigate the impact of agent token quantity in the ESAA module by varying their proportion relative to the sequence length. As shown in Table~\ref{tab:xr-agent}, removing ESAA improves FPS but degrades lip synchronization. Introducing a small proportion achieves the highest FPS, while increasing to 0.5\% yields the best balance between synchronization accuracy and efficiency. Further increasing to 1.0\% offers slight improvements in LMD but reduces overall speed, indicating a trade-off between quality and real-time performance.

\section{Conclusion}

We propose EGSTalker, an efficient and high-fidelity audio-driven talking head generation framework based on 3D Gaussian Splatting (3DGS). By integrating Static Gaussian Initialization and Efficient Gaussian Deformation, EGSTalker achieves comparable generation quality to state-of-the-art 3DGS-based methods while significantly improving inference speed. The Efficient Spatial-Audio Attention (ESAA) module enables effective fusion of spatial and audio features, enhancing synchronization and expressiveness. Experimental results validate the effectiveness of our approach in rendering quality, lip-sync accuracy, and efficiency. Future work will focus on further optimizing inference speed and extending the model to multi-speaker and multimodal scenarios.




\section*{ACKNOWLEDGMENT}

This research was financially supported by the National Natural Science Foundation of China (Grants Nos. 62463029, 62472368, and 62303259) and the Natural Science Foundation of Tianjin (Grant No. 24JCQNJC00910).


\addtolength{\textheight}{-12cm}   

\bibliographystyle{IEEEtran}   
\bibliography{ref}

\end{document}